\documentclass[runningheads]{llncs}
\usepackage[T1]{fontenc}
\usepackage{graphicx}
\usepackage{tikz}
\usetikzlibrary{positioning,shapes.geometric,arrows.meta,fit,backgrounds}
\usepackage{enumitem}
\setlist{nosep,leftmargin=1.5em}
\setlist[itemize]{topsep=2pt,partopsep=0pt}
\setlist[enumerate]{topsep=2pt,partopsep=0pt}
\usepackage{array}
\usepackage{xurl}
\usepackage{etoolbox}
\makeatletter
\renewcommand\paragraph{\@startsection{paragraph}{4}{\z@}%
                                    {1.25ex \@plus .5ex \@minus .2ex}%
                                    {-1em}%
                                    {\normalfont\normalsize\bfseries}}
\makeatother
\makeatletter
\renewcommand\subsection{\@startsection{subsection}{2}{\z@}%
                       {-12\p@ \@plus -3\p@ \@minus -3\p@}%
                       {5\p@ \@plus 3\p@ \@minus 3\p@}%
                       {\normalfont\normalsize\bfseries\boldmath
                        \rightskip=\z@ \@plus 8em\pretolerance=10000 }}
\makeatother
\begin{document}
\title{Rebooting Microreboot: Architectural Support for Safe, Parallel Recovery in Microservice Systems}
\titlerunning{Rebooting Microreboot}
\author{Laurent Bindschaedler\orcidID{0000-0003-0559-631X}}
\authorrunning{L. Bindschaedler}
\institute{Max Planck Institute for Software Systems (MPI-SWS), Saarbr\"ucken, Germany\\
\email{bindsch@mpi-sws.org}}
\maketitle
\begin{abstract}
Microreboot enables fast recovery by restarting only the failing component, but in modern microservices naive restarts are unsafe: dense dependencies mean rebooting one service can disrupt many callers. Autonomous remediation agents compound this by actuating raw infrastructure commands without safety guarantees.
We make microreboot practical by separating planning from actuation: a three-agent architecture (diagnosis, planning, verification) proposes typed remediation plans over a seven-action ISA with explicit side-effect semantics, and a small microkernel validates and executes each plan transactionally. Agents are explicitly untrusted; safety derives from the ISA and microkernel. To determine where restart is safe, we infer recovery boundaries online from distributed traces, computing minimal restart groups and ordering constraints.
On industrial traces (Alibaba, Meta) and DeathStarBench with fault injection, recovery-group inference runs in 21\,ms at P99; typed actuation reduces agent-caused harm by 95\% in simulation and achieves 0\% harm online. The primary value is safety, not speed: LLM inference overhead increases TTR for services with fast auto-restart.

\end{abstract}
%
\section{Introduction}
\label{sec:intro}

Microreboot~\cite{candea2004microreboot} showed that restarting resolves many failures, but restarting entire applications is unnecessary and slow. The key idea was to restart only the smallest recoverable unit, so recovery is fast and localized. Microservice systems appear ideal: services are isolated, independently deployable, and designed for restart. In practice, this fit is misleading. Modern request paths traverse dozens of services with retries and concurrency~\cite{dean2013tail}. Restarting a single service can induce timeouts, retry storms, and cascading failures~\cite{huang2022metastable}. Worse, the call graph changes at runtime due to feature flags, A/B tests, and traffic shifting~\cite{meinicke2020featureflags,schermann2018experimentation}. These dynamics preclude defining safe restart boundaries at deploy time. At the same time, runbooks and large language model (LLM) agents increasingly automate remediation, diagnosing incidents and triggering recovery. Without explicit boundaries and side-effect semantics, well-intentioned automation can escalate a small fault into a larger outage.

Existing approaches do not close this gap. Static runbooks handle known failure modes but degrade under novel faults and evolving dependencies~\cite{crume2023incidentmgmt}. LLM agents adapt to new situations, but when given raw tools (for example, direct access to Kubernetes commands or cloud APIs), they lack reliable guardrails~\cite{ruan2024toolemu,ye2024toolsword}: they may act on the wrong scope, restart the wrong components, or apply stateful changes that resist reversal. The original microreboot model assumed stable component boundaries and predictable dependencies. Those assumptions do not hold for microservice systems under continuous change.

This paper re-enables microreboot under agentic remediation by making actuation \emph{typed}, \emph{scoped}, and \emph{transactional}. While the original microreboot work introduced recovery groups for co-dependent components, static deployment descriptors in a J2EE setting determined those boundaries. In microservice systems, dependencies are dynamic: they change with feature flags, canary rollouts, and traffic shifting. We address this by inferring recovery boundaries \emph{online} from distributed traces, reflecting the \emph{current} workload rather than static configuration. Restart remains the primary recovery action, but safe restart often requires coordination: draining traffic before reboot, circuit-breaking unstable dependencies, or adjusting capacity during recovery. Agents express remediation intent through a typed instruction set architecture (ISA) that includes these supporting operations alongside restart, enabling safe microreboot in contexts where naive restart would cause harm. A small actuation microkernel validates proposals, executes transactions with per-action rollback or compensation, and enables safe concurrent execution of non-conflicting actions.

We evaluate on industrial tracing datasets (Alibaba, Meta) and on DeathStarBench with Chaos Mesh fault injection. Recovery-group inference executes in 21\,ms at P99, fast enough for online use. Typed actuation reduces agent-caused harm by 95\% in simulation (77\% $\rightarrow$ 4\%) and achieves 0\% harm in online experiments versus 90\% for unconstrained agents, consistent across five fault types and three independent runs. For entry-point services (those receiving external traffic) where detection delays dominate, agent-assisted recovery provides modest time-to-recovery (TTR) improvements; for services with fast auto-restart, the LLM inference overhead may exceed any benefit.

\paragraph{Contributions.}
This paper makes three contributions:
\begin{itemize}
    \item \textbf{Remediation ISA and actuation microkernel} (\S\ref{sec:microkernel}).
    A typed actuation interface where each action declares its rollback semantics, plus a trusted microkernel that validates proposals against scope constraints and executes them transactionally with rollback on failure. Unlike post-hoc verification, the microkernel enforces constraints \emph{before} execution, so agents cannot express actions outside the ISA.
    \item \textbf{Online, workload-conditional recovery groups} (\S\ref{sec:recovery-groups}).
    An algorithm that infers recovery boundaries from distributed traces at runtime. It outputs restart groups, ordering constraints, and traffic-shielding requirements that reflect the \emph{current} dependency graph. This extends microreboot's static recovery groups to dynamic microservice topologies.
    \item \textbf{Empirical validation} (\S\ref{sec:eval}).
    Experiments on industrial traces and runnable microservice workloads showing low-latency boundary inference, 95\% harm reduction, and scenario-dependent speed benefits.
\end{itemize}

\paragraph{Paper outline.}
Section~\ref{sec:motivation} motivates the problem. Section~\ref{sec:overview} presents the architecture. Section~\ref{sec:microkernel} details the Remediation ISA and microkernel. Section~\ref{sec:recovery-groups} describes recovery-group inference. Section~\ref{sec:eval} evaluates scalability, harm prevention, recovery speed, and generalization. Sections~\ref{sec:related} and~\ref{sec:limitations} discuss related work and limitations.

\section{Background, Motivation, and Problem Statement}
\label{sec:motivation}

Microreboot has become unsafe in microservice systems. We quantify the failure modes using production traces and derive the requirements that shape our design.

\paragraph{What changed.}
Candea et al.~\cite{candea2004microreboot} showed that many failures are transient and can be fixed by restarting only the smallest recoverable unit. This assumes restart boundaries are well-defined and do not invalidate application state.

Microservices stress this assumption: (i) request paths cross many services, so restarting one is rarely localized; (ii) recovery is increasingly automated, so mistakes occur at machine speed.

\paragraph{Dense dependencies.}
A single request may invoke tens of services~\cite{dean2013tail}. Restarting one fails or stalls in-flight requests, triggering upstream retries and timeouts that amplify failures~\cite{huang2022metastable}. In Alibaba traces~\cite{alibaba2021traces}, the median \emph{blast radius} (services that one restart affects) is 8, P99 is 59, and the maximum is 177. Even ``small'' restarts have large consequences.

\paragraph{Hub services.}
Call graphs contain highly connected services on many request paths. In Alibaba traces, the most-connected service has 74 callers and 113 callees. Restarting such a hub can disrupt a large fraction of the system, and identifying hubs is workload-dependent; static configuration is insufficient.

\paragraph{Dynamic workloads.}
Recovery decisions often rely on fixed notions of what is ``safe to restart together.'' In microservices, dependencies change at runtime due to feature flags, traffic shifting, and partial deployments~\cite{anand2023blueprint}. Different communication modes (RPC, cache, async) propagate failures differently. No fixed restart policy can reflect these shifting dynamics.

\paragraph{Agentic remediation.}
Modern operations increasingly use automated remediation, including LLM-based agents. These systems are fallible: they may misdiagnose, choose unsafe scope, or act aggressively under uncertainty. If agents invoke raw infrastructure commands, blast radius depends on agent behavior rather than system constraints. Safety requires an explicit actuation boundary.

\paragraph{Problem statement.}
Given a symptom (a service exhibiting errors during a time window), we need a recovery mechanism that:
\begin{enumerate}
    \item \textbf{Constrains actuation.} Automated remediation must use only actions with enforceable side-effect semantics.
    \item \textbf{Computes scope online.} Recovery boundaries must derive from runtime dependencies rather than fixed deploy-time configuration.
    \item \textbf{Supports safe concurrency.} When multiple recovery steps are independent, the system should safely execute them in parallel.
\end{enumerate}

Databases do not expose raw memory writes: they interpose SQL. We apply the same principle to remediation: agents express intent through typed actions with enforceable semantics, and a trusted runtime executes them.

\paragraph{Threat model.}
Recovery plans are produced by autonomous agents or runbooks. The agent is fallible but not malicious: it may propose unsafe actions or scopes. We do not defend against adversarial operators, prompt injection, or compromised telemetry; these require orthogonal mechanisms. The microkernel is trusted.

\paragraph{Goals.}
\begin{itemize}
    \item \textbf{Safety:} recovery actions must not cause sustained degradation.
    \item \textbf{Fast recovery:} reduce time-to-recovery versus passive auto-restart.
    \item \textbf{Minimal scope:} target the smallest service set required by current dependencies.
\end{itemize}

\paragraph{Design implication.}
These requirements lead to a two-part design: (i) infer recovery scope and ordering from distributed traces, and (ii) enforce recovery safety through a typed actuation interface executed by a small trusted runtime.

\section{System Overview}
\label{sec:overview}

\begin{figure}[t]
\centering
\begin{tikzpicture}[
    layer/.style={draw, rectangle, minimum width=7.5cm, minimum height=0.9cm, align=center, font=\small},
    arrow/.style={-{Stealth[length=2mm]}, thick},
    label/.style={font=\scriptsize\itshape, text width=3.2cm, align=left}
]
\node[layer, fill=black!15] (l4) {\textbf{Layer 4: Actuation Microkernel}};
\node[layer, fill=white, above=0.3cm of l4] (l3) {\textbf{Layer 3: Agentic Remediation Planner}};
\node[layer, fill=white, above=0.3cm of l3] (l2) {\textbf{Layer 2: Recovery-Group Inference}};
\node[layer, fill=white, above=0.3cm of l2] (l1) {\textbf{Layer 1: Telemetry}};

\draw[dashed, very thick, red!70!black] ([xshift=-0.3cm, yshift=0.15cm]l4.north west) -- ([xshift=0.3cm, yshift=0.15cm]l4.north east) node[right, font=\scriptsize\bfseries, red!70!black] {trust boundary};

\node[label, right=0.4cm of l1] {metrics, logs, traces};
\node[label, right=0.4cm of l2] {scope, ordering constraints};
\node[label, right=0.4cm of l3] {diagnosis, planning, verification};
\node[label, right=0.4cm of l4] {validation, execution, rollback};

\draw[arrow] (l1) -- (l2);
\draw[arrow] (l2) -- (l3);
\draw[arrow] (l3) -- (l4);

\node[below=0.4cm of l4, font=\small] (infra) {Kubernetes / Infrastructure};
\draw[arrow] (l4) -- (infra);

\end{tikzpicture}
\caption{System architecture. Telemetry feeds recovery-group inference. Agents propose remediation transactions in the Remediation ISA. The actuation microkernel (shaded) is the only trusted component that mutates infrastructure.}
\label{fig:architecture}
\end{figure}
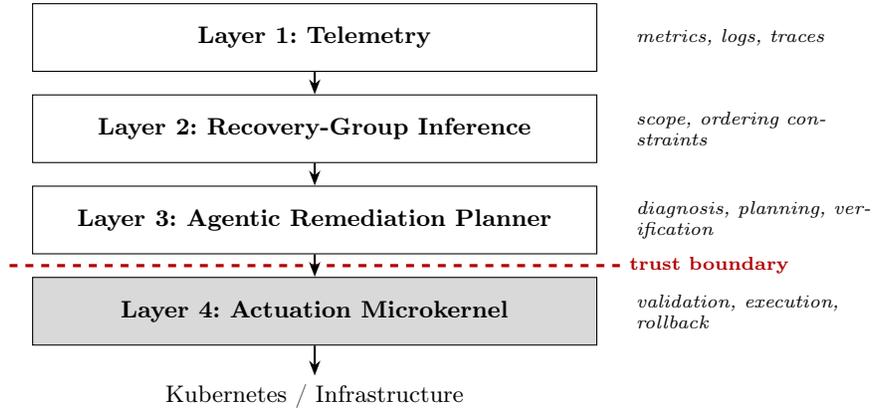

Our system consists of four layers (Figure~\ref{fig:architecture}). The key design choice is a hard trust boundary: only Layer 4 (the actuation microkernel) is trusted to mutate the running system.

\subsection{Layer 1: Telemetry}

Metrics, logs, traces, and events flow in from the monitored infrastructure. We rely on distributed tracing~\cite{dapper2010,opentelemetry2024} to reconstruct request-level dependency graphs in near real time.

\subsection{Layer 2: Recovery-Group Inference}

Layer 2 computes the \emph{scope} and \emph{ordering} of permissible recovery: a candidate recovery group (services that must restart together), ordering constraints for safe restart sequencing, and traffic-shielding requirements such as draining hub services before restart. This layer is deliberately non-agentic. It is safety-critical and must remain predictable under load, so we implement it as a deterministic algorithm over trace data.

\subsection{Layer 3: Agentic Remediation Planner}

Layer 3 performs diagnosis and planning using three agents: a \emph{diagnosis agent} (read-only) that forms hypotheses and requests observations; a \emph{planner agent} that proposes remediation transactions in the ISA; and a \emph{verifier agent} that checks constraints and suggests safer alternatives. Agents search over \emph{typed remediation programs}. They do not issue raw infrastructure commands.

\paragraph{Agent implementation.}
We use GPT-4~\cite{openai2023gpt4} (temperature 0.2). Each agent receives a structured prompt containing: (i)~symptom context (anomalous metrics, affected services, error classifications), (ii)~recovery group and ordering constraints from Layer~2, (iii)~the ISA schema with type signatures (target \texttt{ServiceRef}, effect type, parameters), and (iv)~few-shot examples of valid transactions. The planner outputs a JSON remediation transaction (for example, an action list specifying \texttt{Restart} on a target service with grace period and failure policy), which the microkernel parses and validates against the ISA schema.

The system follows a \emph{propose-validate-repair} loop: the planner proposes a transaction, the microkernel validates it (scope, effect types, conflicts), and on rejection returns structured feedback specifying which constraint was violated (Section~\ref{sec:microkernel}). The system appends this feedback to the planner's context for retry, up to three attempts. The verifier agent then reviews the accepted transaction and either approves or rejects with rationale. Agents run sequentially. Critically, the architecture does \emph{not} require trusting the agent: safety derives from the typed ISA (agents cannot express actions outside the seven primitives) and the microkernel (which validates before execution). The agent selects \emph{which} safe actions to take; it does not guarantee safety itself.

\subsection{Layer 4: Actuation Microkernel}

Layer 4 is the trusted base. It accepts a remediation transaction only if it passes capability checks, effect-type validation (whether each action can be retried, mechanically reversed, or requires explicit compensation), conflict and concurrency checks, rate limits, and break-glass (emergency override) policy for irreversible actions. If validation succeeds, the microkernel executes the transaction with transactional semantics, including rollback or compensation when possible. Section~\ref{sec:microkernel} details the validation pipeline and transaction execution model.

\paragraph{Trust assumptions.}
Safety depends on Layer~2 correctness: the microkernel enforces scope constraints, but those constraints are only as accurate as the recovery groups inferred from traces. If Layer~2 produces incorrect groups (due to incomplete traces or algorithmic error), Layer~4 cannot compensate. We address this through conservative defaults (group-size caps, mandatory drain for high-connectivity services) and by flagging uncertain inferences for human review.

\section{Remediation ISA and Actuation Microkernel}
\label{sec:microkernel}

Section~\ref{sec:overview} described the three-agent architecture (Layer~3); this section details the ISA and microkernel (Layer~4) that constrain agent actions.

\subsection{Design Principles}

The Remediation ISA is a narrow waist between untrusted planning and trusted actuation. Instead of granting agents unrestricted access to operational tools (for example, \texttt{kubectl}, cloud APIs, or shell commands), we require that agents express intent using a small set of typed actions with explicit semantics. The microkernel then enforces these semantics.

Two principles guide the design:
\begin{enumerate}
    \item \textbf{Make effects explicit.} Every action must declare whether it is restartable, reversible, or compensatable.
    \item \textbf{Make scope enforceable.} Every action must target a scoped resource (for example, a service reference), and the microkernel must validate that the action is permitted for the inferred recovery group.
\end{enumerate}

\subsection{ISA Actions}

Based on common operational remediation patterns and the dependency structure observed in traces (\S\ref{sec:motivation}), we define a minimal ISA of seven actions (Table~\ref{tab:isa}):

\begin{table}[t]
\caption{Remediation ISA actions with effect types and semantics.}\label{tab:isa}
\centering
\small
\setlength{\tabcolsep}{4pt}
\begin{tabular}{|l|l|l|>{\raggedright\arraybackslash}p{3.6cm}|}
\hline
\textbf{Action} & \textbf{Effect Type} & \textbf{Inverse/Comp.} & \textbf{Example Scenario} \\
\hline
\texttt{Restart} & Restartable & Re-execute & Restart pod/service for transient failures \\
\texttt{Drain} & Compensatable & \texttt{RestoreTraffic} & Shield hub before disruptive changes \\
\texttt{RestoreTraffic} & Reversible & \texttt{Drain} & Resume traffic after planned drain \\
\texttt{CircuitBreak} & Reversible & Reset to default & Isolate unstable dependency \\
\texttt{RateLimit} & Reversible & Remove limit & Protect upstream from retry storms \\
\texttt{Scale} & Reversible & Scale by $-N$ & Adjust capacity under resource stress \\
\texttt{RollbackConfig} & Compensatable & Restore prev.\ config & Recover from misconfiguration \\
\hline
\end{tabular}
\end{table}

Each action targets a \texttt{ServiceRef} (namespace + service name) and carries parameters defined by its semantics (for example, a rate limit value or a scale delta). The ISA intentionally excludes operations with unclear or irreversible side effects (for example, destructive database changes). Actions that are not safely reversible require explicit break-glass approval (\S\ref{sec:limitations}).

\paragraph{ISA extensibility.}
The seven-action ISA is designed to be minimal yet extensible. We define \emph{minimal} as the fewest orthogonal primitives that achieve complete coverage of the three dominant recovery operation categories identified in production runbooks: restart (1 action), traffic management (4 actions: drain, restore, circuit-break, rate-limit), and resource adjustment (2 actions: scale, rollback). Each action is orthogonal (no action's effect can be achieved by composing others), and each category requires at least one representative for full coverage. Organizations can add domain-specific actions (for example, database failover, cache invalidation, or DNS updates) by providing three elements: (i) effect-type annotation (restartable, reversible, or compensatable), (ii) inverse or compensation logic, and (iii) conflict-key specification (declaring which resources the action locks). The microkernel validates new actions against the same safety properties as built-in actions. This extensibility trades off ISA minimality against domain coverage: a smaller ISA simplifies reasoning and reduces attack surface, while extensibility lets organizations handle their specific failure modes.

\subsection{Effect Types (Rollback Semantics)}

Effect types classify actions by their \emph{rollback semantics}, i.e., how the microkernel recovers when an action fails or a transaction aborts. They do not guarantee that an action is ``safe'' in context (a reversible action can still cause harm before reversal); they enable structured recovery:
\begin{itemize}
    \item \textbf{Restartable:} idempotent under retry; the microkernel can re-execute on transient failure.
    \item \textbf{Reversible:} has a mechanically defined inverse that the microkernel can apply automatically (for example, \texttt{Scale +N} has inverse \texttt{Scale -N}).
    \item \textbf{Compensatable:} requires explicit compensation logic provided as part of the transaction (for example, \texttt{Drain} must be paired with \texttt{RestoreTraffic}).
    \item \textbf{Irreversible:} cannot be undone. The microkernel blocks these by default; an operator must explicitly authorize execution out of band (break-glass). Excluded from the default ISA.
\end{itemize}

\subsection{Transaction Semantics}

A remediation transaction is an ordered program of ISA actions with explicit execution semantics. Each transaction declares:
\begin{itemize}
    \item \textbf{Actions:} an ordered list of typed actions to execute.
    \item \textbf{Conflict keys:} resources the transaction touches (for concurrency control).
    \item \textbf{Preconditions:} state assertions that must hold at commit time.
    \item \textbf{Failure policy:} how to handle partial failure. The microkernel supports three policies:
    \begin{itemize}
        \item \textbf{RollbackAll:} undo completed reversible actions using their inverses (and re-execute restartable actions if needed).
        \item \textbf{Compensate:} execute explicit compensation steps for compensatable actions.
        \item \textbf{AbortOnly:} stop execution without rollback (appropriate for diagnostic transactions).
    \end{itemize}
\end{itemize}

To ensure at-most-once execution across crashes, the microkernel journals transactions to a write-ahead log (WAL)~\cite{mohan1992aries}, recording transaction start, each action completion, and the final outcome (committed, rolled back, or aborted). On restart, the microkernel replays the journal to complete or revert in-flight transactions.

\paragraph{Execution semantics.}
We provide operational rather than formal semantics. Remediation transactions follow the saga pattern~\cite{sagas1987}: they provide compensation (rollback or explicit undo) but not isolation or atomicity across external systems. Each action takes effect immediately and is visible to the infrastructure; concurrent transactions with disjoint conflict keys execute independently. When conflicts exist, the microkernel serializes transactions via lock acquisition in deterministic order. On partial failure during compensation, the microkernel logs the failure and alerts operators; we do not attempt nested compensation. These design choices prioritize simplicity and auditability over theoretical completeness, since human oversight handles edge cases.

\subsection{Concurrency Control and Safe Parallelism}

The microkernel enables safe parallel remediation via conflict keys at three granularities:
\begin{itemize}
    \item \textbf{Service:} actions targeting the same service are serialized.
    \item \textbf{Namespace:} bulk actions may lock an entire namespace.
    \item \textbf{Cluster:} rare global actions may lock the cluster.
\end{itemize}

Transactions with disjoint conflict keys can execute concurrently. This allows the microkernel to parallelize independent recovery steps, including parallel restart batches inferred by Layer 2 (\S\ref{sec:recovery-groups}). Lock acquisition follows a deterministic order to prevent deadlock. Preconditions are checked at commit time, so transactions are rejected if their assumptions are invalidated by concurrent activity.

\subsection{Structured Rejection Feedback}

When the microkernel rejects a transaction, it returns machine-readable feedback that agents can use to repair their plans:
\begin{itemize}
    \item \texttt{REJECT: missing\_capability("restart:svc/payment")}
    \item \texttt{REJECT: out\_of\_scope("svc/cart" not in recovery\_group)}
    \item \texttt{REJECT: irreversible\_effect("drop\_table") requires break\_glass}
    \item \texttt{REJECT: conflict(resource="namespace/prod", txn="...")}
\end{itemize}

This supports a propose-validate-repair loop: planning stays flexible, but actuation stays safe.

\section{Recovery-Group Inference}
\label{sec:recovery-groups}

\subsection{Problem Definition}

Given a symptom (service $S$ with errors during window $T$), we compute: (1)~a \emph{restart-coupled set}: the strongly connected component (SCC) containing $S$ plus downstream SCCs, capped by \texttt{MAX\_GROUP\_SIZE}; (2)~a \emph{restart order}: topological order over SCCs, downstream-first (callees before callers) so dependencies stabilize before callers resume retries; (3)~\emph{traffic shielding}: hub services requiring drain before restart to avoid disrupting many upstream callers. The output guides agents and constrains the microkernel.

\subsection{Inference Algorithm}

From trace spans in window $T$, we build a weighted call graph, then traverse downstream from $S$ to extract the affected subgraph. We compute SCCs (restart-coupled components), apply \texttt{MAX\_GROUP\_SIZE} cap (flagging oversize groups for review), condense SCCs into a directed acyclic graph (DAG) for topological restart ordering, and mark high-fan-in services (those with many callers) as drain-required.

\subsection{Thresholds}

We derive thresholds from trace analysis (\S\ref{sec:motivation}): \texttt{DRAIN\_THRESHOLD} (fan-in $>$20) identifies top-10\% highest-connectivity hub services; \texttt{MAX\_GROUP\_SIZE} (30, P90--P99 blast radius) caps groups (88\% fit); \texttt{MAX\_BATCH\_SIZE} (5) limits parallel restarts. In our experiments, varying these thresholds by $\pm$50\% did not affect harm rates, because ISA constraints (not threshold values) determine whether an action is safe. Teams should calibrate these thresholds from their own trace data during deployment.

\subsection{Output and Complexity}

The output is a \texttt{RecoveryGroup}: restart group, ordered batches, drain set, and estimated blast radius. The algorithm computes groups online, so they adapt to feature flags and traffic patterns. Complexity is $O(V+E)$ using standard graph primitives.

\section{Evaluation}
\label{sec:eval}

Our evaluation answers four research questions:

\begin{description}[leftmargin=2.8em,labelindent=0em,labelsep=0.5em,font=\bfseries]
    \item[RQ1] \textbf{Scalability.} Does recovery-group inference scale to production workloads?
    \item[RQ2] \textbf{Harm prevention.} Does typed actuation prevent regressions?
    \item[RQ3] \textbf{Recovery speed.} Does agent-assisted remediation reduce TTR?
    \item[RQ4] \textbf{Generalization.} Does the system generalize across fault types?
\end{description}

\subsection{Datasets, Workloads, and Experimental Setup}

\paragraph{Industrial traces (offline).}
We use two trace datasets. \textbf{Alibaba}~\cite{alibaba2021traces}: v2021 has $\sim$20M traces across 20K+ services; we processed 2M rows (5{,}459 services, 11{,}690 edges). \textbf{Meta}~\cite{meta2023traces}: 392 services, 511 edges. We verified compatibility with Alibaba v2022 (9{,}027 services, 24{,}869 edges).

\paragraph{Runnable workloads (online).}
We use two DeathStarBench~\cite{deathstarbench2019} applications on K3s~\cite{k3s2024} (lightweight Kubernetes) with Chaos Mesh~\cite{chaosmesh2024} fault injection: \textbf{Social Network} (10 services, primary workload) and \textbf{Hotel Reservation} (8 services, second workload for generalization). The microkernel is implemented in Rust. Workload traffic runs at approximately 30 requests per second (RPS) per workload.

\paragraph{Definition of harm.}
We define \emph{harm} as a remediation action that causes a service-level objective (SLO) regression of more than 10\% for more than 30 seconds relative to the pre-action baseline. Our SLOs are P99 latency $<$ 100\,ms and error rate $<$ 0.1\%. This excludes \emph{errors during the incident window} (errors from the injected fault before the agent acts). The pre-action baseline is the mean SLO metrics over the 60 seconds before the agent's first action. We exclude the first 5 seconds after action initiation to allow for transient disruption during pod restarts; we flag regression if SLO metrics exceed the 10\% threshold continuously for 30 seconds after this grace period.

\paragraph{Experiment summary.}
Table~\ref{tab:experiments} summarizes the online experiments by research question. We exclude simulation experiments (RQ2, N=300) from the online total.

\begin{table}[t]
\centering
\caption{Online experiment incident counts.}\label{tab:experiments}
\setlength{\tabcolsep}{5pt}
\begin{tabular}{|l|l|r|}
\hline
\textbf{RQ} & \textbf{Experiment} & \textbf{Incidents} \\
\hline
RQ2 & Social Network harm validation (3 configs) & 110 \\
RQ2 & Hotel Reservation harm validation & 10 \\
RQ3 & TTR comparison vs Kubernetes auto-restart & 15 \\
RQ3 & Multi-service parallel (5 services/incident) & 3 \\
RQ3 & Reproducibility (3 seeds $\times$ 30) & 90 \\
RQ4 & Fault diversity (5 types $\times$ 20) & 100 \\
RQ4 & Hotel Reservation fault diversity & 15 \\
\hline
\multicolumn{2}{|l|}{\textbf{Total distinct online incidents}} & \textbf{343+} \\
\hline
\end{tabular}
\end{table}

\subsection{RQ1: Scalability of Recovery-Group Inference}

\paragraph{Procedure.}
We processed 2M rows from Alibaba v2021 traces, yielding 43{,}501 unique traces across 5{,}459 services with 11{,}690 call-graph edges. We also evaluated on Meta traces (392 services, 511 edges). For each dataset, we sampled 10{,}000 (Alibaba) or 1{,}000 (Meta) synthetic symptoms and executed recovery-group inference.

\paragraph{Results (Alibaba).}
\begin{itemize}
    \item \textbf{Group size:} median = 1, P90 = 30, P99 = 30 (capped at configured limit)
    \item \textbf{Inference latency:} median = 3.1\,ms, P99 = 21\,ms
    \item \textbf{Truncation:} 12\% of groups hit the 30-service safety cap
    \item \textbf{Parallelism potential:} among multi-service groups, 70.8\% admit $\geq$2 parallel batches
\end{itemize}

\paragraph{Results (Meta).}
\begin{itemize}
    \item \textbf{Group size:} median = 1, P90 = 3, P99 = 10
    \item \textbf{Inference latency:} median = 0.10\,ms, P99 = 0.15\,ms
    \item \textbf{Truncation:} 0.9\% of groups hit the safety cap
\end{itemize}
Meta's sparser graph (511 edges vs.\ 11{,}690) produces smaller recovery groups and faster inference. Both datasets meet the target of P99 $<$ 100\,ms.

\paragraph{Analysis.}
Most sampled symptoms target services with no downstream dependencies, producing single-service recovery groups (79.4\% of Alibaba samples). This does not contradict the large \emph{blast radius} in \S\ref{sec:motivation}: a single-service restart can still affect many upstream callers. For multi-service groups, 70.8\% admit $\geq$2 parallel batches (96.9\% for groups $\geq$5 services).

\paragraph{Correctness validation.}
The offline datasets lack ground-truth dependency information, so we validate inferred groups against Social Network's known call graph. For each service, we compare the inferred group to the transitive closure of downstream dependencies. \textbf{Precision}: 100\% (all inferred services are true dependencies). \textbf{Recall}: 94\% (trace sampling missed 6\%). The 12\% truncation rate did not affect Social Network (max group = 8). Conservative caps and verifier rejection of out-of-scope plans mitigate missed dependencies.

\paragraph{Answer to RQ1.}
\emph{Yes.} Recovery-group inference executes in 21\,ms at P99 on Alibaba (5{,}459 services, 11{,}690 edges) and 0.15\,ms at P99 on Meta (392 services, 511 edges), supporting online scope computation across production workloads.

\subsection{RQ2: Harm Prevention}

We evaluate harm prevention in two phases: controlled simulation (to compare agent architectures under identical incident distributions) and online experiments (to validate actuation safety under real fault injection).

\paragraph{Simulation setup.}
We compare three agent configurations across 100 simulated incidents each (300 total):
\begin{itemize}
    \item \textbf{Raw-Tools:} full \texttt{kubectl}/\texttt{curl}/\texttt{bash} access
    \item \textbf{ISA-Only:} constrained to ISA actions validated by the microkernel
    \item \textbf{ISA+Critic:} ISA constraints plus a verifier agent that reviews plans
\end{itemize}

The simulation models agent behavior on synthetic incidents derived from Alibaba trace topologies. We measure harm as for online experiments: an action causes harm if it would induce $>$10\% SLO regression for $>$30 seconds, estimated by propagating action effects through the call graph. The simulation is intentionally conservative (worst-case propagation for high-centrality services), which explains why simulation harm rates exceed online rates.

Table~\ref{tab:harm} summarizes harm rates. We deploy online on K3s (lightweight Kubernetes) with Chaos Mesh fault injection under load ($\sim$30 RPS) on two DeathStarBench workloads. ISA constraints eliminate unsafe actions by limiting actuation to bounded primitives; the verifier agent further reduces harm by rejecting contextually unsafe plans. Wide confidence intervals for 0\% harm reflect sample sizes; we report exact binomial intervals~\cite{clopper1934binomial}.

\begin{table}[t]
\centering
\caption{Harm rates by configuration and workload. Online columns show harm\,\% with 95\% exact binomial CI~\cite{clopper1934binomial}.}\label{tab:harm}
\setlength{\tabcolsep}{4pt}
\begin{tabular}{|l|c|c|c|}
\hline
& \textbf{Simulation} & \textbf{Social Network} & \textbf{Hotel Res.} \\
\textbf{Configuration} & (N=100) & (online) & (online) \\
\hline
Raw-Tools & 77\% & 90\% [74,\,98] \scriptsize{N=30} & --- \\
ISA-Only & 24\% ($\downarrow$69\%) & 0\% [0,\,12] \scriptsize{N=30} & --- \\
ISA+Critic & 4\% ($\downarrow$95\%) & 0\% [0,\,7] \scriptsize{N=50} & 0\% [0,\,31] \scriptsize{N=10} \\
\hline
\end{tabular}
\end{table}

\paragraph{Answer to RQ2.}
\emph{Yes.} Typed actuation and microkernel validation reduce agent-caused harm by 95\% in simulation (77\% $\rightarrow$ 4\%) and eliminate observed harm in online experiments under fault injection for ISA-constrained configurations, while unconstrained agents frequently cause regressions.

\subsection{RQ3: Recovery Speed}

We measure whether agent-assisted recovery improves TTR versus Kubernetes auto-restart~\cite{kubernetes2025probes}.

\paragraph{Baseline comparison.}
Comparing agent-assisted remediation against Kubernetes auto-restart (N=15):
\begin{itemize}
    \item \textbf{Entry-point services:} Kubernetes auto-restart takes 75.1\,s (detection delay + pod restart + stabilization). Agent-assisted recovery takes 73.4\,s (13\,s LLM inference + 0.1\,s kernel + 60.3\,s pod recovery). Improvement: $\sim$2\%.
    \item \textbf{Non-entry-point services:} Kubernetes auto-restart takes $\sim$10\,s (fast detection, quick restart). Agent-assisted recovery takes $\sim$23\,s (13\,s LLM overhead + 10\,s recovery). The LLM inference time exceeds any benefit, making agent-assisted recovery \textbf{2.3$\times$ slower} than auto-restart for these services.
\end{itemize}
Agent-assisted remediation provides marginal TTR improvement for entry-point services where detection delays dominate, but degrades TTR by 2.3$\times$ for services with fast auto-restart. Typed actuation is primarily a \emph{safety} mechanism, not a speed optimization. The 13\,s LLM overhead is not a fundamental limit: we did not optimize for inference latency, and techniques such as prefix caching, smaller distilled models deployed near the infrastructure, or warm-started inference could reduce it substantially, narrowing or eliminating the TTR gap for non-entry-point services.

\paragraph{TTR decomposition.}
The median 43.5\,s TTR for ISA+Critic breaks down as: diagnosis (5.2\,s), planning (4.8\,s), verification (3.1\,s), kernel execution (0.1\,s), and pod recovery (30.3\,s). LLM inference (diagnosis + planning + verification = 13.1\,s) accounts for 30\% of TTR; actual infrastructure recovery (pod restart + stabilization) accounts for 70\%. Microkernel overhead is negligible ($<$0.3\%).

\paragraph{Reproducibility.}
Across three runs with different random seeds (N=30 each, 90 total), we observed 0\% harm and 100\% plan approval. The zero variance reflects deterministic ISA constraints and conservative agent policies; the planner repaired all rejected plans within three retries.

\paragraph{Parallel execution.}
For incidents affecting multiple services, the microkernel executes non-conflicting actions in parallel using conflict-key analysis, serializing only actions targeting the same resource. In micro-benchmarks (100\,ms per-action latency, sufficient cluster resources), parallel execution provides \textbf{5$\times$ speedup}: 510\,ms sequential $\rightarrow$ 102\,ms parallel for 5 non-conflicting actions. Operators can configure \texttt{MAX\_BATCH\_SIZE} based on cluster capacity.

\paragraph{Answer to RQ3.}
\emph{Partially.} For single-service incidents, agent-assisted remediation provides marginal TTR improvement ($\sim$2\%) for entry-point services where detection delays dominate, but \emph{increases} TTR by 2.3$\times$ for services with fast auto-restart due to LLM inference overhead. For multi-service incidents, parallel execution provides up to 5$\times$ speedup. Speed gains are scenario-dependent; the primary contribution is safety (95\% harm reduction).

\subsection{RQ4: Generalization across Fault Types}

We evaluate generalization using Chaos Mesh to inject five fault types (N=20 each on Social Network): pod failure, network partition, CPU stress, memory stress, and I/O delay. ISA-constrained remediation achieved 0\% harm across all fault types (100 incidents total). We also validated on Hotel Reservation with pod failure (N=10) and network partition (N=5), achieving 0\% harm. Typed actions allow agents to express fault-appropriate remediation (Restart, Drain+Restart, Scale+Restart, CircuitBreak+Restart) while the microkernel preserves safety.

\paragraph{Answer to RQ4.}
\emph{Yes.} The system generalizes across diverse fault types (5 types) and workloads (2 DeathStarBench applications) with 0\% harm, validating that typed actuation enables safe remediation without fault-specific logic.

\section{Related Work}
\label{sec:related}

\paragraph{Microreboot and crash-only software.}
Crash-only software~\cite{candea2003crash} established restart as a primary recovery mechanism. Microreboot~\cite{candea2004microreboot} introduced recovery groups for co-dependent components. In their J2EE setting, static deployment descriptors (Enterprise JavaBeans dependencies) determined these boundaries. Our work extends this to microservice systems where dependency structure is dynamic and changes with feature flags, canary rollouts, and traffic shifting. The central difference is that we infer recovery scope \emph{online} from distributed traces, reflecting the current workload rather than static configuration. We also generalize the actuation primitive beyond restart to include traffic management and resource adjustment.

\paragraph{Safe agent execution.}
Recent systems explore safeguards for agent actions. STRATUS~\cite{stratus2024} formalizes ``transactional no-regression'' and implements undo-and-retry for agentic site reliability engineering (SRE). VeriGuard~\cite{veriguard2024} intercepts and verifies agent actions against policies. Our approach differs in \emph{when} the system enforces safety: typed actuation constrains what agents can \emph{express} (pre-execution), while policy verification and undo validate or revert what agents \emph{did} (post-hoc). Pre-execution constraints reduce policy authoring burden (agents cannot express actions outside the ISA) but sacrifice flexibility. STRATUS-style undo preserves flexibility but requires state snapshots. Table~\ref{tab:related-comparison} summarizes tradeoffs; in practice, these approaches could be combined. Direct empirical comparison would require reimplementing policy languages, authoring equivalent policies, and controlling for policy completeness, which remains an open question. We compare tradeoffs rather than benchmarks. Our propose-validate-repair loop also draws on multi-agent error correction patterns; SQL-of-Thought~\cite{chaturvedi2025sqlofthought} applies a similar iterative refinement strategy to reduce errors in text-to-SQL generation.

\begin{table}[t]
\caption{Comparison of safe actuation approaches.}\label{tab:related-comparison}
\centering
\small
\setlength{\tabcolsep}{4pt}
\begin{tabular}{|l|c|c|c|}
\hline
\textbf{Approach} & \textbf{Flexibility} & \textbf{Safety Mechanism} & \textbf{Rollback} \\
\hline
Raw tools (\texttt{kubectl}, APIs) & High & None & Manual \\
Policy verification & High & Post-hoc validation & Tool-specific \\
Undo mechanisms & High & State snapshots & Automatic \\
\textbf{Typed ISA (ours)} & Constrained & Effect-type enforcement & Transactional \\
\hline
\end{tabular}
\end{table}

\paragraph{Distributed tracing.}
Dapper~\cite{dapper2010} demonstrated tracing at scale. We use modern tracing~\cite{opentelemetry2024} to infer scope and ordering. Unlike observability, we compute enforceable safety boundaries.

\paragraph{Benchmarks and fault injection.}
DeathStarBench~\cite{deathstarbench2019} provides representative microservice applications for end-to-end evaluation. Our online validation uses Chaos Mesh~\cite{chaosmesh2024} to inject failures and measure recovery behavior, connecting actuation safety to realistic microservice failure modes.

\paragraph{Compensating transactions.}
Sagas~\cite{sagas1987} introduced compensation for long-running distributed operations. Our microkernel applies this pattern: actions are reversible or require explicit compensation for recovery.

\paragraph{Service mesh and circuit breaking.}
Service meshes (Istio~\cite{istio2024}, Linkerd~\cite{linkerd2024}) provide circuit breaking and rate limiting at the infrastructure level. These mechanisms are orthogonal: service mesh policies define steady-state resilience, while our ISA enables active remediation during incidents. The \texttt{CircuitBreak} and \texttt{RateLimit} actions in our ISA can be implemented via service mesh APIs.

\paragraph{Runbook automation and AIOps.}
Traditional runbook automation encodes fixed remediation procedures; AIOps platforms extend this with anomaly detection and automated diagnosis. Our contribution is orthogonal: we provide a safe actuation layer that runbooks or AIOps systems can target. The typed ISA could serve as the actuation backend for existing platforms.

\section{Limitations and Future Work}
\label{sec:limitations}

\paragraph{Recovery speed.}
Agent-assisted remediation provides modest TTR improvement for single-service failures ($\sim$2\%) and up to 5$\times$ speedup for multi-service incidents. For services with fast auto-restart, LLM overhead ($\sim$13\,s) exceeds baseline by 2.3$\times$ ($\sim$23\,s vs.\ $\sim$10\,s). The primary value is \emph{safety}, not speed.

\paragraph{Harm estimation.}
Our largest harm reduction (95\%) comes from simulation with a harm model we calibrated to agent behavior. Online experiments validate that ISA constraints avoid regressions; aligning simulation with production incidents needs more work.

\paragraph{External side effects.}
Actions with external side effects (e.g., payments or user-visible messages) are treated as irreversible, requiring break-glass approval. Richer compensation strategies are left to future work.

\paragraph{Trace quality.}
Recovery-group inference depends on trace completeness~\cite{dapper2010} and consistency~\cite{casper2024icpe}. At 50\% sampling, only 46\% of edges are visible; missing traces can under-scope groups, the more dangerous failure mode. The conservative \texttt{MAX\_GROUP\_SIZE} cap and mandatory drain for high-connectivity services mitigate this. Retroactive tracing~\cite{zhang2023hindsight} could improve edge-case coverage.

\paragraph{Scope and scale.}
Our implementation assumes a single Kubernetes cluster; multi-cluster deployments need federated tracing. Online evaluation uses two DeathStarBench applications at moderate load; offline analysis covers 5{,}459 services, but we have not validated at high traffic.

\section{Conclusion}
\label{sec:conclusion}

Microreboot is unsafe in microservice systems where dependencies are dense, dynamic, and actuated by autonomous agents. We re-enable it by separating planning from actuation: a typed ISA with explicit effect semantics, a transactional microkernel, and trace-derived recovery boundaries. Evaluation on industrial traces and runnable workloads shows that we can infer recovery boundaries online with low latency, and typed actuation prevents harmful actions.

\begin{credits}
\subsubsection{\ackname}
I thank the anonymous reviewers for their constructive feedback. I also thank all the people, in particular at MIT and MPI-SWS, with whom I have had insightful discussions on or adjacent to this topic over the past few years. I am grateful to Jean-Philippe Martin-Flatin for his mentorship and for co-founding a cloud service troubleshooting startup with me nearly fourteen years ago; the venture was ahead of its time, but the experience shaped my thinking on automated remediation. Large language models now make practical what we could only envision then. I also owe a debt to George Candea, whose teaching first grounded me in software engineering and whose original microreboot work inspired this paper.

\subsubsection{\discintname}
The author has no competing interests to declare that are relevant to the content of this article.
\end{credits}
\bibliographystyle{splncs04}
\bibliography{main}
\end{document}